\documentclass[12pt,journal,onecolumn]{IEEEtran}

\usepackage{cite,authblk,rotating,setspace,amsmath,amssymb,theorem,epstopdf}
\usepackage[caption=false, font=footnotesize]{subfig}

\newtheorem{theorem}{Theorem}

\newenvironment{Proof}[1]{\medskip\par\noindent{\bf Proof:\,}\,#1}{{\mbox{\,$\blacksquare$}\par}}

\allowdisplaybreaks

\begin{document}

\title{Timely Distributed Computation with Stragglers\thanks{This work was supported by NSF Grants CCF 17-13977 and ECCS 18-07348.}}

\author{Baturalp Buyukates \qquad Sennur Ulukus\\
	\normalsize Department of Electrical and Computer Engineering\\
	\normalsize University of Maryland, College Park, MD 20742\\
	\normalsize  \emph{baturalp@umd.edu} \qquad \emph{ulukus@umd.edu}}

 %\normalsize {\it ypwei@umd.edu  \qquad \it kbanawan@umd.edu} \qquad {\it ulukus@umd.edu}}         

\maketitle

\vspace*{-1.0cm}

\begin{abstract}	
	We consider a status update system in which the update packets need to be processed to extract the embedded useful information. The source node sends the acquired information to a computation unit (CU) which consists of a master node and $n$ worker nodes. The master node distributes the received computation task to the worker nodes. Upon computation, the master node aggregates the results and sends them back to the source node to keep it \emph{updated}. We investigate the age performance of uncoded and coded (repetition coded, MDS coded, and multi-message MDS (MM-MDS) coded) schemes in the presence of stragglers under i.i.d.~exponential transmission delays and i.i.d~shifted exponential computation times. We show that asymptotically MM-MDS coded scheme outperforms the other schemes. Furthermore, we characterize the optimal codes such that the average age is minimized.
\end{abstract}

\section{Introduction}

Age of information metric has been widely studied as a timeliness metric in real-time systems producing time-sensitive information. In these systems, time-critical data are collected and sent to the monitor node(s). Thus, most of the work on age of information focuses on the queueing-theoretic framework under various arrival and service profiles \cite{Kaul12a,Costa16,Yates12,Najm17,Inoue18b,Soysal18, Zhong17a, Buyukates18, Buyukates19, Maatouk19, Tripathi19, Bedewy19, Wang19b}. Another line of research studies the age from optimization, scheduling and energy harvesting perspectives \cite{Kadota18b, Zhou18b, Bastopcu18, Bastopcu19, Buyukates18c, Buyukates19b, Bacinoglu15,Wu18,Arafa18a,Arafa18b,Farazi18, Ornee19, Leng19, Stamatakis19, Elmagid19, Zheng19}.

In many real-time monitoring applications including autonomous driving, surveillance systems and predictive maintenance, time-sensitive data that are collected by sensors or mobile devices require processing to extract the embedded information. However, these devices cannot perform heavy computations due to battery related issues or their limited computational capabilities. These type of status update packets that require computation are called \emph{computation-intensive messages}. References that are most closely related to our work are \cite{Kuang19, Gong19, Arafa19, Song19, Zhong19, Zou19b, Arafa19b} which study queueing, packet management and scheduling in such status update systems. Common to all these works is the fact that they consider a single computation server. 

\begin{figure}[t]
	\centering  \includegraphics[width=0.7\columnwidth]{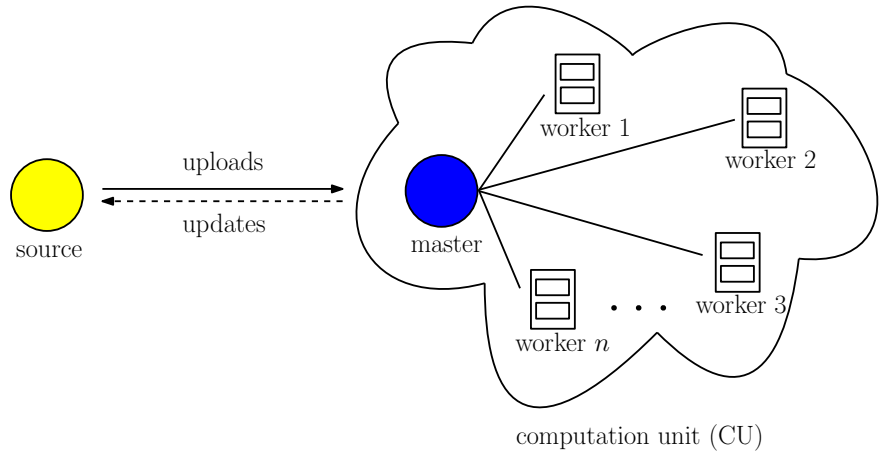}
	%\vspace{-1em}
	\caption{System model with a single source node and a computation unit (CU) that consists of a master node and $n$ identical worker nodes.}
	\label{fig:model}
	%\vspace{-0.5cm}
\end{figure}

In this work, we consider a system in which there is a source node which uploads computation-intensive time-critical status updates to a computation unit (CU) which consists of a single master node and $n$ worker nodes that perform the computations (see Fig.~\ref{fig:model}). We assume that the required computation on the update is a linear operation such as large matrix multiplication. This brings up the concept of computation distribution among the worker nodes. Computation distribution and scheduling problem has been extensively studied particularly in the context of machine learning with a focus on completion time and straggler threshold analysis \cite{Lee18, Tandon17, Das18, Dutta18, Amiri19, Ozfatura18, Ozfatura19, Ozfatura19b, Aktas19, Yang19, Duffy19}. 

Inspired by the recent distributed computation literature, we investigate uncoded and coded computation distribution algorithms to combat the stragglers, i.e., nodes that are slower than the average, and to minimize the average age of information in the system. The source node collects time-sensitive data and sends them to the CU for processing over a channel with random transmission delays. Arriving packets at the CU go into service (computation) if the CU is idle by the time of their arrival. Otherwise, they are dropped. Here, the master node distributes the overall computation to $n$ worker nodes using uncoded or coded schemes. Computation time at each worker is random. Once the master node collects sufficiently many results from the worker nodes to decode the computation result, it \emph{updates} the source node. We note that in our model destination node is also the source node. One such application is autonomous driving cars which capture images/videos of the surroundings and send them to a CU for computation. The source node in the mean time adopts a zero-wait policy such that it sends the next update packet once the current one reaches the CU.

We derive the average age for uncoded and coded schemes and show that asymptotically multi-message MDS (MM-MDS) coded scheme outperforms the uncoded, repetition coded and MDS coded schemes. Our results also indicate that given that the source node and the CU implement zero-wait and dropping policies, respectively, when the transmission delays are i.i.d.~exponentials and computation times are i.i.d.~shifted exponentials, for large $n$, minimizing age of information is equivalent to minimizing the computation time. Finally, we find the optimal repetition, MDS and MM-MDS codes that minimize the average age of information.

\section{System Model and Age Metric} \label{model}

We consider a system (see Fig.~\ref{fig:model}), where there is a single source node which sends time-sensitive computation-intensive status updates, i.e., packets that require additional processing to extract the embedded information, to a CU which consists of a single master node and $n$ worker nodes. Worker nodes have statistically identical computing capabilities. From the source node to the CU, update packets experience i.i.d.~exponential transmission delays. Upon successful arrival of a packet to the CU, the master node distributes this computation task to $n$ worker nodes. Here, we use status update packet and computation task interchangeably. Each worker node performs the computation, which is assumed to be a linear operation, and sends the result back to the master node. We note that one such computation task example is large matrix multiplication prevalent in machine learning applications. 

Computation times of the workers are modeled by i.i.d.~shifted exponential random variables. The constant shift makes sure that computation times cannot go below a certain value whereas the exponential part constitutes the tail of the computation time distribution. This is inline with the computation times observed in systems like Google Trace \cite{Aktas19}. 

When the master node receives sufficiently many responses from the worker nodes, it aggregates the results and \emph{updates} the source node. We neglect the transmission delay from the CU back to the source node after computation as the size of the initial packet is in general much larger than the resulting update packet after computation.

Thus, in our model, packets that are able to enter the CU experience two stages: transmission and computation. Random variable $D_j$ denotes the transmission delay of the $j$th update packet and is exponentially distributed with parameter $\lambda$. To model the computation times at the worker nodes, we adopt the model in \cite{Lee18} and assume the existence of a mother runtime distribution. This distribution corresponds to the computation time when the whole computation on the update is performed by a single worker, $X$, and has a shifted exponential distribution with $(c,\mu)$ where $c>0$ is the shift and $\mu$ is the rate which is also the straggling parameter. When the update packet is divided into $m$ subpackets, the computation time of each subpacket has the scaled-down (i.e., sped-up) version of the overall distribution, i.e., shifted exponential with $\left(\frac{c}{m}, m\mu\right)$.

The source node adopts a zero-wait policy in which it sends the next update as soon as the current one reaches the CU. On the other hand, the CU implements a dropping policy in which when busy it neglects any update packets arriving from the source node. Thus, packets sent by the source node can only enter the computation stage if the CU is idle at the time of their arrival. Upon finishing a computation task, the CU immediately sends back the result and waits for the next packet arrival. This idle waiting time is denoted by random variable $Z$ and is exponentially distributed with $\lambda$ because of the memoryless property of the transmission delays $D$. 

To distribute the computation task among the worker nodes upon receiving status update packets, the master node may adopt uncoded or coded distribution algorithms. In the uncoded scheme, the status update packet is divided into $n$ equal subpackets, one for each worker node. However, in this method, the overall computation time is limited by the slowest worker node and thus, it may not be desirable especially when the computations are time-sensitive. To combat these slower straggling worker nodes, the master node can implement coding techniques to introduce redundancy to the computation task so that some straggling nodes can be tolerated. In our model, we analyze repetition and MDS (maximum distance separable) codes. Moreover, to further utilize the fastest worker nodes, we investigate the assignment of multiple MDS coded subpackets to each worker node, i.e., multi-message MDS (MM-MDS). We assume that communications within the CU in between the master node and worker nodes are instantaneous. We analyze the effects of these uncoded and coded schemes on the timeliness of the computations. 

To quantify the timeliness we use the age of information metric. At time $t$ age at the destination node, which is the source node in our model, is a random process $\Delta(t) = t - u(t)$ where $u(t)$ is the time-stamp of the most recent update at the destination node. The metric we use, time averaged age, is 
\begin{align}
\Delta = \lim_{\tau\to\infty} \frac{1}{\tau} \int_{0}^{\tau} \Delta(t) dt,
\end{align}
where $\Delta(t)$ is the instantaneous age as defined above. 

\section{Age of Uncoded and Coded Task Distribution Algorithms}\label{age_analysis}

From the perspective of the CU, we have i.i.d.~exponential interarrivals with $\lambda$. Since a dropping policy is implemented, not every arriving packet actually goes into service at the CU. We denote the packets that find the CU idle and thus go into service as the successful packets. Let $T_{j-1}$ and $T'_{j-1}$ denote the time at which the $j$th successful packet is generated at the source node and is received by the CU, respectively. Random variable $Y$ denotes the update cycle at the CU, time in between two consecutive successful arrivals, and $Y_j = T'_j - T'_{j-1}$. As described in Section~\ref{model}, update cycle $Y_j$ consists of computation (service) time $S_j$ and idle waiting time $Z_j$. We note that $Z_j$, $S_j$ and $D_j$ are mutually independent where $D_j$ denotes the transmission delay experienced by the $j$th successful packet, i.e., $D_j = T'_{j-1} - T_{j-1}$. In our model, the interarrival process at the CU, $D$, and service times $S$ are independent, and sequences $\{D_1, D_2,\dots \}$ and $\{S_1, S_2,\ldots \}$ form i.i.d.~processes.

We observe that $Z$ is stochastically equal to the transmission delay $D$, i.e., interarrival time at the CU, due to the memoryless property of the transmission delay $D$. On the other hand, computation time $S$ changes depending on the task distribution algorithm adopted by the master node. We use order statistics to express the distribution of $S$. We denote the $k$th smallest of $X_1, \dots, X_n$ as $X_{k:n}$. For a shifted exponential random variable $X$ with $(c, \mu)$, we have
\begin{align}
E[X_{k:n}] =& c + \frac{1}{\lambda}(H_n - H_{n-k}), \label{ord1} \\
Var[X_{k:n}] =& \frac{1}{\lambda^2}(G_{n} - G_{n-k}), \label{ord2}
\end{align}
where $H_n = \sum_{j=1}^{n} \frac{1}{j}$ and $G_{n} = \sum_{j=1}^{n} \frac{1}{j^2}$. Using (\ref{ord1}) and (\ref{ord2}),
\begin{align}
E[X_{k:n}^2] =& \left(c + \frac{1}{\lambda}(H_n \!- H_{n-k}) \right)^2 \!+\! \frac{1}{\lambda^2}\left(G_{n} \!- G_{n-k} \right). \label{ord3}
\end{align}

We note for future reference that when $n$ is large and $k$ is linear in $n$, i.e., $k = \alpha n$, for $0<\alpha<1$, the variance of the $k$th order statistic of the shifted exponential random variable, $X_{k:n}$, shown in (\ref{ord2}), becomes negligibly small and tends to $0$ as $n$ increases because both $G_n$ and $G_{n-k}$ sequences converge to $\frac{\pi^2}{6}$. Thus, for large $n$, an ordered sequence of $n$ shifted exponential random variables essentially becomes a deterministic sequence such that the $k$th realization takes the mean value given by (\ref{ord1}). This observation will be useful in designing age optimal codes.

\begin{figure}[t]
	\centering  \includegraphics[width=0.6\columnwidth]{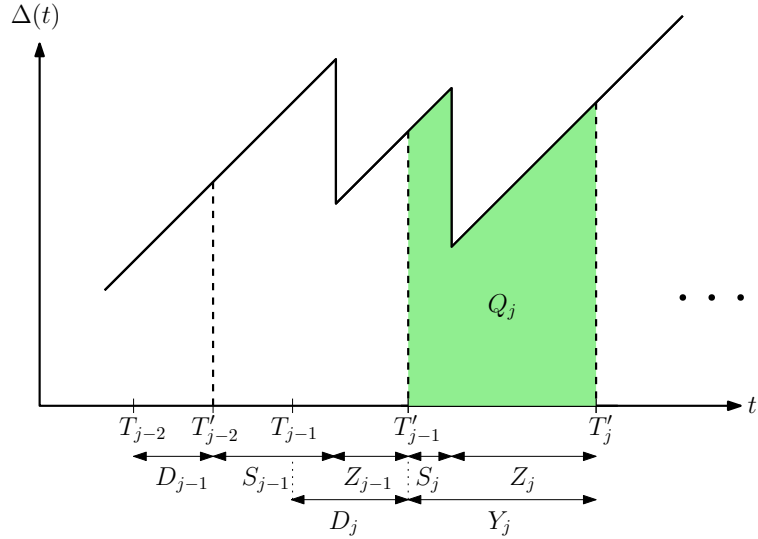}
	%\vspace{-1em}
	\caption{Sample age evolution $\Delta{(t)}$ at the destination (source) node. Successful updates are indexed by $j$. The $j$th successful update leaves the source node at $T_{j-1}$ and arrives at the CU at $T'_{j-1}$. Update cycle at the CU is the time in between two successive arrivals and is equal to $Y_j = S_j + Z_j = T'_{j} - T'_{j-1}$.}
	\label{fig:ageEvol}
	%\vspace{-0.5cm}
\end{figure}

Our model here resembles the $M/G/1/1$ queue with blocking model analyzed in \cite{Najm17} with one difference:  here arriving packets at the CU have experienced a transmission delay. Therefore, they have aged by $D$ by the time of their arrival at the CU. Noting this difference and performing a similar graphical analysis using Fig.~\ref{fig:ageEvol}, we find 
\begin{align}
E[Q] =& E[D](E[S] + E[Z]) +  \frac{1}{2} E[(S+Z)^2] + E[S](E[S]+E[Z]), \label{EQ} \\ 
E[Y] =& E[S] + E[Z], \label{EL}
\end{align}
where $Q$ denotes the shaded area and $Y$ is its length. Then, assuming ergodicity, long term time average age is
\begin{align}
\Delta = \frac{E[Q]}{E[Y]} = E[D] + E[S] + \frac{E[Y^2]}{2E[Y]}, \label{avg_age}
\end{align}
where $Y = S + Z$ as noted above. The first term in (\ref{avg_age}) reflects the fact that arriving packets at the CU have aged on average by $E[D]$.\footnote{We note that, in a possibly more intuitive setting, if the source node sends the next update upon receiving the previous computed update back from the CU, unlike the current model in which the source node sends the next update as soon as the current one reaches the CU, the average age expression in (\ref{avg_age}) would stay the same as $D$ and $Z$ are stochastically identical.} Our goal is to minimize the average age given in (\ref{avg_age}) by adjusting computation (service) time $S$ at the CU through different task distribution algorithms. 

\subsection{Uncoded Scheme}

In the uncoded scheme, the master node divides the received status update packet into $n$ subpackets, one for each worker node. Using the mother runtime distribution detailed in Section~\ref{model}, we see that in the uncoded scheme local computation time at each worker $\tilde{X}$ follows a shifted exponential distribution with parameters $\left(\frac{c}{n}, n\mu\right)$. This sped-up distribution highlights the fact that each worker node performs a part of the overall computation. Thus, in order to reveal the information in the received update packet, the master node needs to collect the results from all $n$ worker nodes. Hence, the computation time is $S = \tilde{X}_{n:n}$. Calculating the moments of $S$ using (\ref{ord1})-(\ref{ord3}) and substituting them into ($\ref{avg_age}$) we find the average age when the uncoded scheme is utilized, $\Delta_{unc}$, as
\begin{align}
\Delta_{unc} =& \frac{1}{\lambda} + \frac{c}{n} + \frac{H_n}{n\mu} + \frac{\left(\frac{c}{n} + \frac{H_n}{n\mu}\right)^2 + \frac{G_n}{n^2 \mu^2} + \frac{2}{\lambda} \left(\frac{c}{n} + \frac{H_n}{n\mu}\right) + \frac{2}{\lambda^2}}{2\left(\frac{c}{n} + \frac{H_n}{n\mu} + \frac{1}{\lambda}\right)}. \label{age_uncoded}
\end{align}

The following theorem states the asymptotic average age performance of the uncoded scheme as the number of worker nodes $n$ increases.

\begin{theorem} \label{thm1}
	With i.i.d.~exponential transmission delays and i.i.d.~shifted exponential computation times at each worker, the average age of the uncoded distribution scheme for large $n$ is $\frac{2}{\lambda} + O\left( \frac{\log n}{n}\right)$.
\end{theorem}

The proof of Theorem~\ref{thm1} follows from the fact that for large $n$, we have $H_n \approx \log n$ and $ G_n \approx \frac{\pi^2}{6}$. The constant $\frac{2}{\lambda}$ in the result reflects the sum of $E[D]$ = $\frac{1}{\lambda}$, which is the expected delay packets experience on the way from the source to the CU, and $E[Z] = \frac{1}{\lambda}$, which is the expected waiting time for a new packet at the CU when it is idle. The  $O\left( \frac{\log n}{n}\right)$ term in the result reflects the fact that the average age decreases with $n$, the number of worker nodes at the CU. The main downside of the uncoded scheme is the fact that it is prone to large delays due to straggling nodes as the master node needs all of the computation results to extract the useful information from the status update packet. Therefore, if some servers are much slower than the rest, service time of the update packet increases significantly. To cope with these straggling worker nodes, redundant computation tasks may be created via coding. In what follows we analyze the effects of repetition coded, MDS coded and MM-MDS coded schemes on the average age. 

\subsection{Repetition Coded Scheme} 

We consider an $\frac{n}{k}$-repetition code where the packet is divided into $k$ equal sized subpackets where $k \leq n$ and each subpacket is repeated $\frac{n}{k}$ times. In other words, each subpacket has $\frac{n}{k}$ replicas and the master node needs to collect $k$ distinct results from $n$ worker nodes. Thus, each worker node has a shifted exponential computation time distribution, $\tilde{X}$, with parameters $\left( \frac{c}{k}, k\mu\right)$. Since there are $\frac{n}{k}$ workers for each of the $k$ subpackets the computation time of each subpacket is the minimum among these $\frac{n}{k}$ i.i.d.~random variables which is denoted by $\bar{X} = \tilde{X}_{1:\frac{n}{k}}$. Since the minimum of shifted exponentials is also a shifted exponential with the same shift but scaled straggling parameter, $\bar{X}$ follows a shifted exponential distribution with $\left(\frac{c}{k}, n\mu\right)$. Since we need $k$ distinct results, the overall computation time in this case is $S=\bar{X}_{k:k}$. Using (\ref{avg_age}) along with the moments of order statistics given in (\ref{ord1})-(\ref{ord3}), we find the average age of the repetition coded scheme, $\Delta_{rep}$, as 
\begin{align}
\Delta_{rep} =& \frac{1}{\lambda} + \frac{c}{k} + \frac{H_k}{n\mu} + \frac{\left(\frac{c}{k} + \frac{H_k}{n\mu}\right)^2 + \frac{G_k}{n^2 \mu^2} + \frac{2}{\lambda} \left(\frac{c}{k} + \frac{H_k}{n\mu}\right) + \frac{2}{\lambda^2}}{2\left(\frac{c}{k} + \frac{H_k}{n\mu} + \frac{1}{\lambda}\right)}. \label{age_rep}
\end{align}

The following theorem states the asymptotic average age performance of the repetition coded scheme as $n$ increases.

\begin{theorem} \label{thm2}
	With i.i.d.~exponential transmission delays and i.i.d.~shifted exponential computation times at each worker, the average age of the $\frac{n}{k}$-repetition coded scheme for large $n$ with $k = \alpha n$ where $0<\alpha \leq 1$ is $\frac{2}{\lambda} + O\left( \frac{\log n}{n}\right)$.
\end{theorem}

The proof of Theorem~\ref{thm2} follows similarly from that of Theorem~\ref{thm1}. Here, we observe that although a coding scheme is implemented, asymptotically, we achieve the same average age performance as the uncoded scheme. Thus, repetition coded scheme is asymptotically no better than the uncoded scheme. Next, we analyze the performance of the MDS coded schemes.

\subsection{MDS Coded Scheme} \label{MDS}

To implement an $(n,k)$-MDS code where $k < n$, the update packet is first divided into $k$ equal sized subpackets. From these $k$ subpackets a total of $n$ subpackets are created with coding by using $n-k$ redundant subpackets so that the master node can decode the result of the computation as soon as it receives $k$ computation results. Since the overall computation task is divided into $k$ subtasks as in repetition coding, each worker node completes its local task in $\tilde{X}$ which is a shifted exponential with $(\frac{c}{k}, k\mu)$. The computation time for the overall task is, however, $S = \tilde{X}_{k:n}$. Using this along with (\ref{ord1})-(\ref{ord3}) in (\ref{avg_age}), we find the average age when an $(n,k)$-MDS code is implemented, $\Delta_{mds}$, as
\begin{align}
\Delta_{mds} =& \frac{1}{\lambda} + \frac{c}{k} + \frac{H_n-H_{n-k}}{k\mu} + 	\frac{\left(\frac{c}{k} + \frac{H_n-H_{n-k}}{k\mu}\right)^2 + \frac{G_n-G_{n-k}}{k^2 \mu^2} + \frac{2}{\lambda} \left(\frac{c}{k} + \frac{H_n-H_{n-k}}{k\mu}\right) + \frac{2}{\lambda^2} }{2\left(\frac{c}{k} + \frac{H_n-H_{n-k}}{k\mu} + \frac{1}{\lambda}\right)}.  \label{age_MDS}
\end{align}

The following theorem gives the asymptotic average age performance of the MDS coded scheme for large $n$.

\begin{theorem} \label{thm3}
	With i.i.d. exponential transmission delays and i.i.d. shifted exponential computation times at each worker, the average age of the $(n,k)$-MDS coded scheme for large $n$ with $k=\alpha n$ where $0<\alpha<1$ is $\frac{2}{\lambda} + O\left(\frac{1}{n}\right)$.
\end{theorem}

The proof of Theorem~\ref{thm3} follows similarly to that of Theorem~\ref{thm1} by noting that $H_{n-k} \approx \log (n-k)$. Thus, when $k = \alpha n$, we get $H_n - H_{n-k} = -\log(1-\alpha)$ which is independent of $n$. Further, $ G_n - G_{n-k} = G_n - G_{(1-\alpha)n}$ tends to $0$ as $n$ increases. With these, the result follows. 

We observe that the average age in Theorem~\ref{thm3} has a $O\left(\frac{1}{n}\right)$ term as opposed to $O\left(\frac{\log n}{n}\right)$ terms in Theorems~\ref{thm1} and~\ref{thm2}. Thus, for large $n$, MDS coded scheme outperforms repetition coded and uncoded schemes in terms of average age performance. Up to now, we have investigated uncoded and coded schemes in which each worker node is assigned one subtask to compute. Although we achieve better performance in combating the straggling nodes through coding, there is still room for improvement. In all these schemes, the fastest worker nodes which finish their computations earlier stay idle. To utilize them even more, we can assign multiple subtasks to each worker node. With multiple assignments to each worker node we can utilize partial straggling worker nodes, which are the ones that cannot finish all tasks that are assigned to them but still return some partial results. In the next subsection, we consider the performance of MDS coded scheme when each worker is given multiple subtasks to compute.  

\subsection{Multi-message MDS (MM-MDS) Coded Scheme}

In multi-message MDS coded scheme, each worker node is assigned $\ell$ subpackets to compute, where $\ell$ denotes the load of each worker node. That is, each worker node has a job queue of size $\ell$ in each update cycle. Thus, we implement an $(n\ell, k)$-MDS code. For this, the overall update packet is divided into $k$ subtasks where $k < n\ell$ and from these subtasks $n\ell - k$ redundant subtasks are generated such that the master node only needs to receive $k$ computation results to extract the embedded information in the status update received from the source node.  Unlike regular MDS coded scheme in which each worker has one subtask to compute, in this scheme faster workers can perform multiple computations to aid the overall computation time. Hence, we utilize partial stragglers, also called non-persistent stragglers \cite{Ozfatura18}, i.e., worker nodes that finish some portion of the subtasks that are assigned to them. 

In line with the mother computation distribution model presented in Section~\ref{model}, computation time of a subtask at each worker, $\tilde{X}$, has a shifted exponential distribution with $\left(\frac{c}{k}, k\mu \right)$. Following the model in \cite{Ozfatura18}, we assume that the duration of each computation performed by a worker during an update cycle is identical. In other words, if a worker finishes $m$ of the $\ell$ subtasks during an update cycle, duration of each computation is identical which is sampled from a shifted exponential with parameters $\left(\frac{c}{k}, k\mu \right)$, i.e., $m\tilde{X}$. Therefore, the time it takes for a worker node to perform $m$ computations is also a shifted exponential with $(\frac{mc}{k}, \frac{k \mu}{m})$. It remains to determine $E[S]$ and $E[S^2]$ to calculate the average age in this setting by using (\ref{avg_age}). 

\begin{figure}[t]
	\centering  \includegraphics[width=0.75\columnwidth]{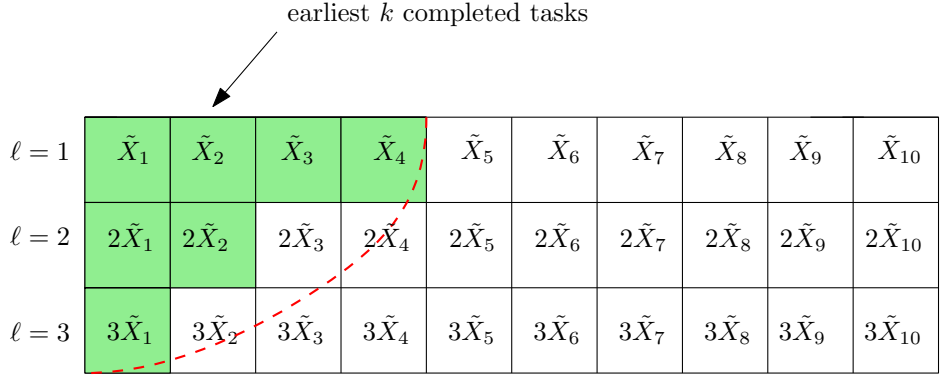}
	%\vspace{-1em}
	\caption{The earliest $k$ computed tasks for $n=10$, $\ell =3 $, and $k=7$.}
	\label{fig:earliestk}
	%\vspace{-0.5cm}
\end{figure}

In what follows, the $m$th level refers to the set of subtasks that are located in the $m$th position in each worker's job queue. In other words, the $m$th level includes a total of $n$ subtasks which are performed in the $m$th position by the corresponding worker nodes upon completion of their first $m-1$ subtasks. We note that in the uncoded, repetition coded, and MDS coded schemes there is only one level as $\ell=1$ in those schemes. Let $k_m$ denote the number of subtasks computed in the $m$th level upon completion of the overall task at the CU during an update cycle.  We have $\sum_{m=1}^{\ell} k_m = k$ since exactly $k$ subtasks need to be performed to finish the overall computation. Fig.~\ref{fig:earliestk} shows an example for $n=10$, $k=7$, and $\ell=3$. Here, each column represents the computation times of $\ell$ subtasks that a worker node is assigned and row $m$ represents the computation times of the $m$th level subtasks. Without loss of generality, we order level one, i.e., $\tilde{X}_1$ in Fig.~\ref{fig:earliestk} is the smallest computation time of a level $1$ subtask and $\tilde{X}_{10}$ is the largest one. Correspondingly, all other levels are ordered as well. Hence, column $i$ in Fig.~\ref{fig:earliestk} in fact shows the computation times of the $i$th fastest worker node, where $i=1, \ldots, n$. In this example, we observe that by the time the earliest $k=7$ computations are finished, the fastest worker completed three subtasks, the second fastest worker completed two subtasks, the third and fourth fastest workers completed one subtask each, and the remaining six workers completed zero subtasks. That is, $4$ first level, $2$ second level and $1$ third level subtasks are computed. Thus, we have $k_1 = 4$, $k_2=2$ and $k_3 =1$.

For simplicity and ease of exposition, consider the case where $\ell = 2$. Assume that by the time a total of $k$ computations are performed $k_1 < k$ computations from the first level are finished. Then, we have $k_2 = k - k_1$. When $n$ is large, the time it takes to compute $k_1$ first level subtasks, $\tilde{X}_{k_1:n}$ is equal to $E[\tilde{X}_{k_1:n}]$ due to the vanishing variance property of order statistics of shifted exponentials for large $n$, discussed after (\ref{ord3}). Similarly, $k_2$ second level subtasks are performed in $E[2\tilde{X}_{k_2:n}]$ units of time. At this point, there can be two possible scenarios: In the first scenario, $k_2$ second level computations are finished before the $k_1$th first level subtask is finished, i.e.,
\begin{align}
E[2\tilde{X}_{k_2:n}] \leq  E[\tilde{X}_{k_1:n}] < E[2\tilde{X}_{k_2+1:n}]. \label{k1_identity}
\end{align} 
We note that second inequality in (\ref{k1_identity}) holds since the $k_1$th first level task needs to be performed before the $k_2+1$th second level task to be included in the earliest $k$ subtasks.
In this case, the computation duration, i.e., service time, is $S = \tilde{X}_{k_1:n}$. On the other hand, in the second scenario, the $k_2$th second level subtask is finished after the $k_1$st first level subtask but before the subtask $k_1+1$ of the first level, i.e.,
\begin{align}
E[\tilde{X}_{k_1:n}] \leq E[2\tilde{X}_{k_2:n}]  < E[\tilde{X}_{k_1+1:n}]. \label{k1_identity_2}
\end{align} 
In this case, we have $S = 2\tilde{X}_{k_2:n}$. When $k_1$ and $k_2$ are linear in $n$, i.e., $k_1 = \alpha_1 n$ and $k_2 = \alpha_2 n$, with $0< \alpha_1<1$ and $0 < \alpha_2<1$, for large $n$, we have
\begin{align}
\alpha_1 = \frac{k_1}{n} &\approx \frac{k_1 +1}{n}, \\ 
\alpha_2 = \frac{k_2}{n} &\approx \frac{k_2 +1}{n}, 
\end{align}
which imply that the bounds in (\ref{k1_identity}) and (\ref{k1_identity_2}) meet as $n$ gets large. Thus, by the sandwich theorem $E[\tilde{X}_{k_1:n}] = E[2\tilde{X}_{k_2:n}]$ is satisfied for large $n$ in either scenario. By using (\ref{ord1}), for large $n$ when both $k_1$ and $k_2$ are linear in $n$ as defined above we have 
\begin{align}
E[\tilde{X}_{k_1:n}] &= \frac{c}{k} + \frac{H_n - H_{n-k_1}}{k\mu} \nonumber \\ &=  \frac{c}{2\alpha n} + \frac{1}{2\alpha n\mu}\log \left(\frac{1}{1-\alpha_1}\right), \label{alpha_1}
\end{align} 
and
\begin{align}
	E[2\tilde{X}_{k_2:n}] &= 2\left(\frac{c}{k} + \frac{H_n - H_{n-k_2}}{k\mu}\right) \nonumber \\ &=  \frac{2c}{2\alpha n} + \frac{1}{2\alpha n\mu}\log \left(\frac{1}{1-\alpha_2}\right)^2. \label{alpha_2}
\end{align}
Here, $k = \alpha 2n$ with $0<\alpha<1$ when both $k_1$ and $k_2$ are linear in $n$ as defined above. We note that $k_1 + k_2 = k$ is equivalent to $\alpha_1 + \alpha_2 = 2\alpha$. Equating (\ref{alpha_1}) and (\ref{alpha_2}) yields
\begin{align}
\frac{1}{1-\alpha_1} = e^{\mu c}\frac{1}{(1-\alpha_2)^2}, \label{alpha_1_v2}
\end{align}
where $\alpha_1 + \alpha_2 = 2\alpha$. We note that (\ref{alpha_1_v2}) holds when MM-MDS coded scheme is implemented for $\ell=2$ if $\alpha_2>0$, i.e., $k_2 > 0$. When $\alpha_2=0$, we have $\alpha_1=2\alpha$ directly. That is, in that case we have $k_1 = k$. 

A similar relationship between $\alpha_m$s, equivalently between $k_m$s, holds for the general case with $\ell>2$ as well. When we have $\ell$ levels, we have at most ${\ell \choose 2}$ inequalities like (\ref{k1_identity}) and (\ref{k1_identity_2}) to represent the ordering between the last subtasks of each level that are included in the earliest $k$. For example, when $\ell = 2$ we have one inequality, either (\ref{k1_identity}) or (\ref{k1_identity_2}), to represent the relationship between the $k_1$th first level and the $k_2$th second level subtasks provided that $k_2$ is nonzero. When $\ell>2$, if $k_m>0$ subtasks are finished in level $m$ by the time it takes to receive a total of $k$ computations, completion time of the $k_m$th subtask of the $m$th level satisfies either one of the following inequalities for any other level $\bar{m}$ for which $k_{\bar{m}}>0$:
\begin{align}
E[m\tilde{X}_{k_m:n}] \leq  E[\bar{m}\tilde{X}_{k_{\bar{m}}:n}] < E[m\tilde{X}_{k_{m}+1:n}] \label{km_identity}
\end{align}
which implies that the $k_m$th subtask of the $m$th level is finished earlier than the $k_{\bar{m}}$th subtask of level $\bar{m}$ or
\begin{align}
E[\bar{m}\tilde{X}_{k_{\bar{m}}:n}] \leq E[m\tilde{X}_{k_m:n}]  < E[\bar{m}\tilde{X}_{k_{\bar{m}}+1:n}], \label{km_identity_2}
\end{align}
which implies that the $k_m$th subtask of the $m$th level is finished after the $k_{\bar{m}}$th subtask of level $\bar{m}$ is finished. Upon writing this relationship between every $(m, \bar{m})$ pair, we take $k_m = \alpha_m n$ with $0 < \alpha_m<1$ and proceed similarly to get 
\begin{align}
\frac{1}{(1-\alpha_{m-1})^{m-1}} &= e^{\mu c}\frac{1}{(1-\alpha_m)^m}, \label{cond3} 
%\sum_{m=1}^{\ell} \alpha_m &= \ell \alpha \label{cond4}
\end{align}
with $\sum_{m=1}^{\ell} \alpha_m = \ell \alpha $ similar to (\ref{alpha_1_v2}). We note that if after some level $m >m_0$, none of the level $m$ subtasks are finished, then, $\alpha_m = 0$ for all $m > m_0$ and (\ref{cond3}) holds for all nonzero $\alpha_m$s. With these, by using (\ref{cond3}) and the fact that $\sum_{m=1}^{\ell} \alpha_m = \ell \alpha $, each remaining nonzero $\alpha_m$ and correspondingly each remaining $k_m$ can be determined. As a direct consequence of (\ref{cond3}), we see that the time it takes to receive the earliest $k$ computation results is equivalent to the time it takes to receive $k_m$ from level $m$ such that $k_m=\alpha_m n$ and $\alpha_m$s satisfy (\ref{cond3}) and $\sum_{m=1}^{\ell} \alpha_m = \ell \alpha $. With such $k_m$s, we then have $S = \tilde{X}_{k_1:n}$. Hence, the average age when the MM-MDS coded scheme is implemented with $\ell$ subpackets at each node, $\Delta_{mm-mds}$, can now be computed using (\ref{avg_age}) as follows
\begin{align}
\Delta_{mm-mds} =& \frac{1}{\lambda} + \frac{c}{k} + \frac{H_n-H_{n-k_1}}{k\mu} + \frac{\left(\frac{c}{k} + \frac{H_n-H_{n-k_1}}{k\mu}\right)^2 + \frac{G_n-G_{n-k_1}}{k^2 \mu^2} + \frac{2}{\lambda} \left(\frac{c}{k} + \frac{H_n-H_{n-k_1}}{k\mu}\right) + \frac{2}{\lambda^2}}{2\left(\frac{c}{k} + \frac{H_n-H_{n-k_1}}{k\mu} + \frac{1}{\lambda}\right)}. \label{age_mm_mds}
\end{align}
where $k_m = \alpha_m n$ and $\alpha_m$s satisfy (\ref{cond3}) and $\sum_{m=1}^{\ell} \alpha_m = \ell \alpha $.

The following theorem gives the asymptotic average age performance of the MM-MDS coded scheme for large $n$.
\begin{theorem} \label{thm_mm_mds}
	With i.i.d exponential transmission delays and i.i.d. shifted exponential computation times at each worker, the average age of the MM-MDS coded scheme with load $\ell$, for large $n$ with $k_m = \alpha_m n$ where $0 < \alpha_m < 1$, $m=1,\ldots,\ell$, is $\frac{2}{\lambda} + O\left(\frac{1}{\ell n}\right)$. 
\end{theorem}

To prove Theorem~\ref{thm_mm_mds} we first note that when $k_m = \alpha_m n$ for each level $m$, we have $k = \alpha n\ell$ where $0<\alpha<1$ such that $\sum_{m=1}^{\ell} \alpha_m = \alpha\ell$. With this, the proof follows similarly from that of Theorem~\ref{thm3}. We note that compared to the MDS coded scheme where we have $O\left(\frac{1}{n}\right)$, here in the MM-MDS coded scheme, we have $O\left(\frac{1}{\ell n}\right)$ which reflects $\ell$, the number of subtasks assigned to each worker node. Thus, for large $n$, the best asymptotic performance is achieved when MM-MDS coded scheme is implemented. 

The performance of repetition coded, MDS coded and MM-MDS coded schemes can be optimized through the selection of $k$, which will be the focus of Section~\ref{age_opt}.

\section{Optimizing Age by Parameter Selection} \label{age_opt}

In Section~\ref{age_analysis}, we showed that the age in uncoded, repetition coded, MDS coded, and MM-MDS coded schemes depend on $n$ as $O\left( \frac{\log n}{n}\right)$, $O\left( \frac{\log n}{n}\right)$, $O\left( \frac{1}{n}\right)$, and $O\left( \frac{1}{\ell n}\right)$, respectively, excluding the common constant term $\frac{2}{\lambda}$. In repetition, MDS and MM-MDS coded schemes, we have a parameter $k$ that depends on $n$ linearly as $k=\alpha n \ell$, where $\ell=1$ for repetition and MDS coded schemes, and $\ell >1$ for MM-MDS coded scheme. In this section, we consider the optimization of this parameter $k$, which is equivalent to the optimization of the parameter $\alpha$. This is similar in spirit to the optimization of $k$, correspondingly the optimization of $\alpha$, in \cite{Zhong17a, Buyukates18, Buyukates19, Lee18}. Towards that goal, in order to unify our approach for all three coded schemes (repetition, MDS, and MM-MDS), we first provide the following theorem. This theorem shows that, in our model, age minimization translates into computation (service) time minimization which is not always the case in age optimization problems. 

\begin{theorem} \label{theorem1}
	When the transmission delays are i.i.d. exponentials and computation times at each worker are i.i.d. shifted exponentials under the dropping policy at the CU, for large $n$, minimization of the average age of repetition coded, MDS coded and MM-MDS coded schemes, is equivalent to minimization of the average computation time. 
\end{theorem}

\begin{Proof} 
	In the repetition, MDS and MM-MDS coded schemes, the computation time $S$ is characterized through the selection of $k$. Using the average age expression in (\ref{avg_age}), the minimization problem is
	\begin{align}
	\min\limits_{k} E[D] + &E[S] + \frac{E[(S+Z)^2]}{2E[S+Z]} \nonumber \\
	&=\min\limits_{k} E[D] + E[S] + \frac{E[S^2] + 2E[S]E[Z] + E[Z^2]}{2E[S+Z]} \label{age_approx2} \\ 
	&\approx \min\limits_{k} E[D] + E[S] + \frac{E^2[S] + 2E[S]E[Z] + E[Z^2]}{2(E[S]+E[Z])} \label{age_approx3} \\
	&= \min\limits_{k} \frac{1}{\lambda} + E[S] + \frac{E^2[S] + 2E[S]\frac{1}{\lambda} + \frac{2}{\lambda^2}}{2(E[S]+\frac{1}{\lambda})} \label{age_approx4} \\
	&= \min\limits_{k} \frac{1}{\lambda} + E[S] + \frac{(E[S] + \frac{1}{\lambda})^2 + \frac{1}{\lambda^2}}{2(E[S]+\frac{1}{\lambda})} \label{age_approx5} \\
	&= \min\limits_{k} \frac{3}{2\lambda} + \frac{3}{2}E[S] + \frac{\frac{1}{\lambda^2}}{2E[S] + \frac{2}{\lambda}}, \label{age_approx6}
	\end{align} 
	where (\ref{age_approx2}) follows from the independence of $S$ and $Z$, and (\ref{age_approx3}) follows from the fact that $E[S^2] \approx E^2[S]$ in all of these coding schemes due to the vanishing variance property of order statistics of shifted exponentials for large $n$, discussed after (\ref{ord3}).

	In order to optimize the average age, we need to select the optimal $k$ in the repetition coded, MDS coded and MM-MDS coded schemes in (\ref{age_approx6}). We note that in (\ref{age_approx6}) only $E[S]$ depends on $k$. Although the second term in (\ref{age_approx6}) increases in $E[S]$ and the third term decreases in $E[S]$, overall (\ref{age_approx6}) is monotonically increasing in $E[S]$, as the derivative of (\ref{age_approx6}) with respect to $E[S]$ is nonnegative. Thus, the average age is minimized when $E[S]$ is minimized. That is, the average age minimization is equivalent to the average computation time minimization.  \end{Proof}

For large $n$, average computation time is given, for the repetition coded scheme, by
\begin{align}
	E[S_{rep}] &= \frac{c}{k} + \frac{H_k}{n\mu} =  \frac{c}{k} + \frac{1}{\mu n}\log k \nonumber \\
	&= \frac{c}{\alpha n} + \frac{1}{\mu n}\log (\alpha n) \label{ESrep}, 
\end{align}
for the MDS coded scheme by
\begin{align}
E[S_{mds}] &= \frac{c}{k} + \frac{H_n - H_{n-k}}{k\mu} = \frac{c}{k} + \frac{1}{\mu k}\log \left(\frac{n}{n-k}\right) \nonumber \\ &=\frac{c}{\alpha n} + \frac{1}{\mu \alpha n}\log \left(\frac{1}{1-\alpha}\right) \label{ESmds}, 
\end{align}
and for the MM-MDS coded scheme by
\begin{align}
E[S_{mm-mds}] &= \frac{c}{k} + \frac{H_n - H_{n-k_1}}{k\mu} =  \frac{c}{k} + \frac{1}{\mu k}\log \left(\frac{n}{n-k_1}\right) \nonumber \\ &= \frac{c}{\alpha n \ell} + \frac{1}{\mu \alpha n \ell}\log \left(\frac{1}{1-\alpha_1}\right), \label{ES_mm_mds}
\end{align}   
where in (\ref{ES_mm_mds}) $k_m$s satisfy (\ref{cond3}) and $\sum_{m=1}^{\ell} \alpha_m = \ell \alpha $. 

Reference \cite{Lee18} finds the optimal $k$ for repetition coded and MDS coded schemes when $k$ is linear in $n$ by noting that both (\ref{ESrep}) and (\ref{ESmds}) have unique extreme points as functions of $k$. In  \cite[Lemma 1]{Lee18} the following optimization problem is solved to find the optimal computation time in the repetition coded scheme:
\begin{align}
\min\limits_{k} E[S_{rep}] &= 	\min\limits_{1 \leq k \leq n}  \frac{c}{k} + \frac{1}{\mu n}\log k  \nonumber \\ &= 	\min\limits_{0 < \alpha \leq 1}  \frac{c}{\alpha } + \frac{1}{\mu }\log \alpha \label{opt_rep}
\end{align}
Objective in (\ref{opt_rep}) has a term that increases in $\alpha$ and another term that decreases in $\alpha$. Depending on $\mu$ and $c$ values, there is a unique $\alpha^*$ which is the extremum point
\begin{equation}
\alpha^* = \begin{cases} 
1, & c\mu \geq 1 \\
c\mu,  & c\mu < 1 \label{repopt_alpha}
\end{cases}
\end{equation}
and correspondingly,
\begin{equation}
k^* = \begin{cases} 
n, & c\mu \geq 1 \\
c\mu n,  & c\mu < 1 \label{repopt}
\end{cases}
\end{equation}
for large $n$. Solutions in (\ref{repopt_alpha}) and (\ref{repopt}) are computation time optimum, and also average age optimum from Theorem~\ref{theorem1} for $\frac{n}{k}$-repetition coded scheme. Note that, for $c\mu \geq 1$, the optimal repetition coded scheme is in fact the uncoded scheme. However, when $c\mu < 1$, repetition coded scheme outperforms the uncoded scheme. 

Similarly, to find the optimal computation time in the $(n,k)$-MDS coded scheme, the following optimization problem is solved in \cite[Lemma 2]{Lee18}:
\begin{align}
\min\limits_{k} E[S_{mds}] &= 	\min\limits_{1\leq k < n}  \frac{c}{k} + \frac{1}{\mu k}\log \left(\frac{n}{n-k}\right)  \nonumber \\ &= 	\min\limits_{0 < \alpha < 1}  \frac{c}{\alpha } + \frac{1}{\mu \alpha }\log \left(\frac{1}{1-\alpha}\right),
\end{align}
and it is shown that the optimum $\alpha$ is 
\begin{align}
\alpha^* = 1+ \frac{1}{W_{-1}(-e^{-\mu c -1})},
\end{align}
and correspondingly,
\begin{align}
k^* = \left[ 1+ \frac{1}{W_{-1}(-e^{-\mu c -1})} \right]n,
\end{align}
for large $n$, which is also average age optimal from Theorem~\ref{theorem1}. Here, $W_{-1}(\cdot)$ is the lower branch of Lambert $W$ function. 

In the MM-MDS coded scheme, as in the repetition coded and MDS coded schemes, by selecting the optimal $k$, correspondingly the optimal $\alpha$, the computation time and by Theorem~\ref{theorem1}, the age can be minimized. To do that, we need to solve the following optimization problem: 
%\begin{equation}
\begin{align}
\min_{0 < \alpha, \alpha_1, \cdots, \alpha_{\ell}< 1} \quad &  \frac{c}{\alpha} + \frac{1}{\mu \alpha }\log \left(\frac{1}{1-\alpha_1}\right)\nonumber\\
\textrm{s.t.} \quad \quad &\frac{1}{(1-\alpha_{m-1})^{m-1}} = e^{\mu c}\frac{1}{(1-\alpha_m)^m}, \quad m=2,\ldots,\ell \nonumber\\ 
&\sum_{m=1}^{\ell} \alpha_m = \ell \alpha \label{mm-mds_optprob_v1}
\end{align} 
%\end{equation} 
To give an explicit example, for instance, for $\ell=3$, we need to solve
%\begin{equation}
\begin{align}
\min_{ 0 < \alpha, \alpha_1, \alpha_2, \alpha_3 < 1} \quad & \frac{c}{\alpha } + \frac{1}{\mu \alpha }\log \left(\frac{1}{1-\alpha_1}\right) \nonumber \\
\textrm{s.t.} \quad \quad & \frac{1}{1-\alpha_1} = e^{\mu c} \frac{1}{(1-\alpha_2)^2} \nonumber \\
&\frac{1}{(1-\alpha_2)^2} = e^{\mu c} \frac{1}{(1-\alpha_3)^3}  \nonumber \\
& \alpha_1 + \alpha_2 + \alpha_3 = 3\alpha \label{mm-mds_optprob} 
\end{align}
%\end{equation}
We note that unlike the repetition coded and MDS coded schemes, in the MM-MDS coded scheme, the optimization problem (\ref{mm-mds_optprob_v1}) is more complicated. The optimization in (\ref{mm-mds_optprob_v1}) is over $\alpha$ and all $\alpha_m$s. Here, a closed-form expression for $k^*$, or equivalently $\alpha^*$, is not available unlike the former two cases. We solve the problem in (\ref{mm-mds_optprob_v1}) in the next section using numerical methods.

\section{Numerical Results}

In this section, we provide simple numerical results. 

\begin{figure}
	\centering
	\subfloat[\label{fig:muone}]{   \includegraphics[width=.495\columnwidth]{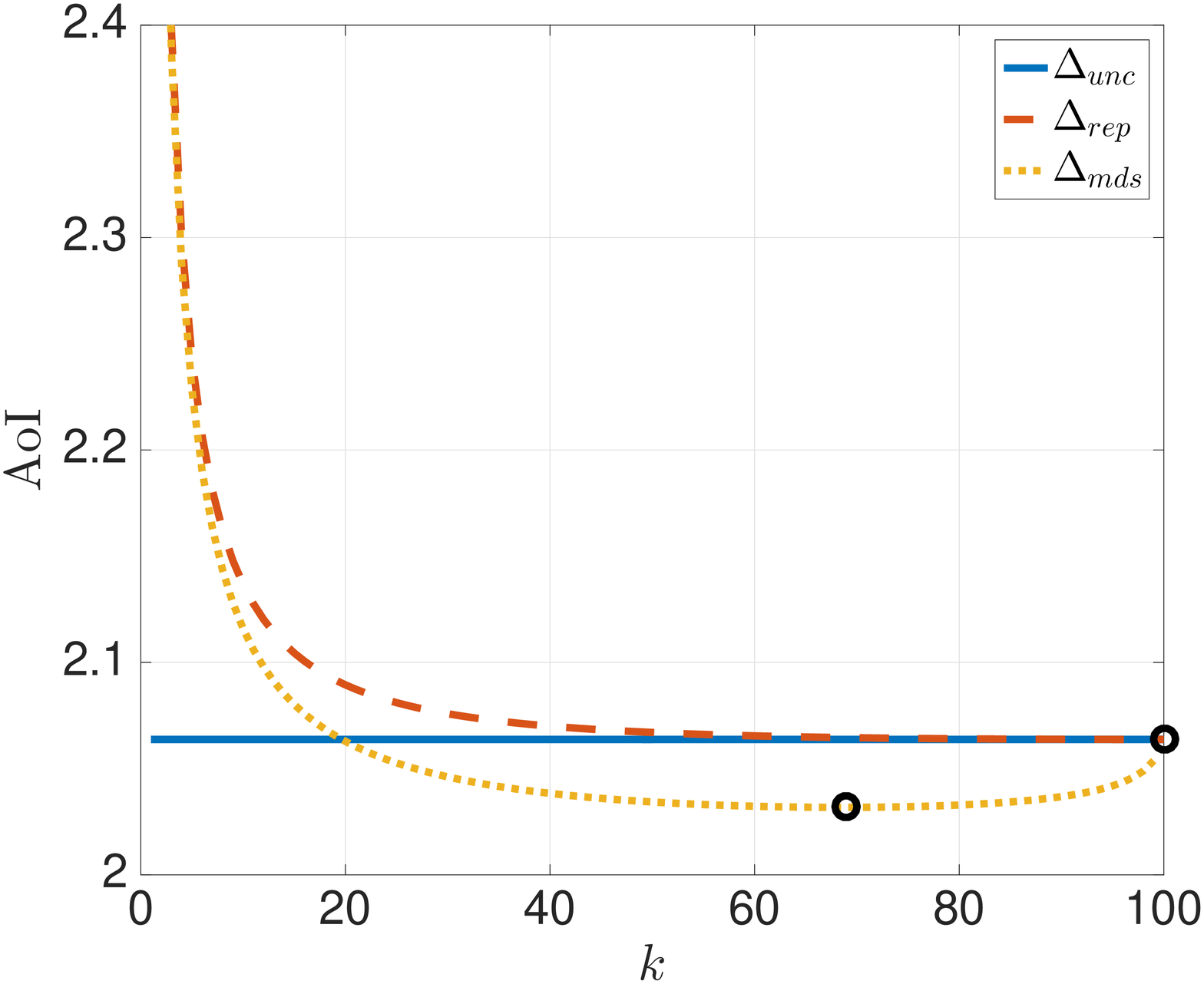}    }%
	\subfloat[\label{fig:muhalf}]{  \includegraphics[width=.495\columnwidth]{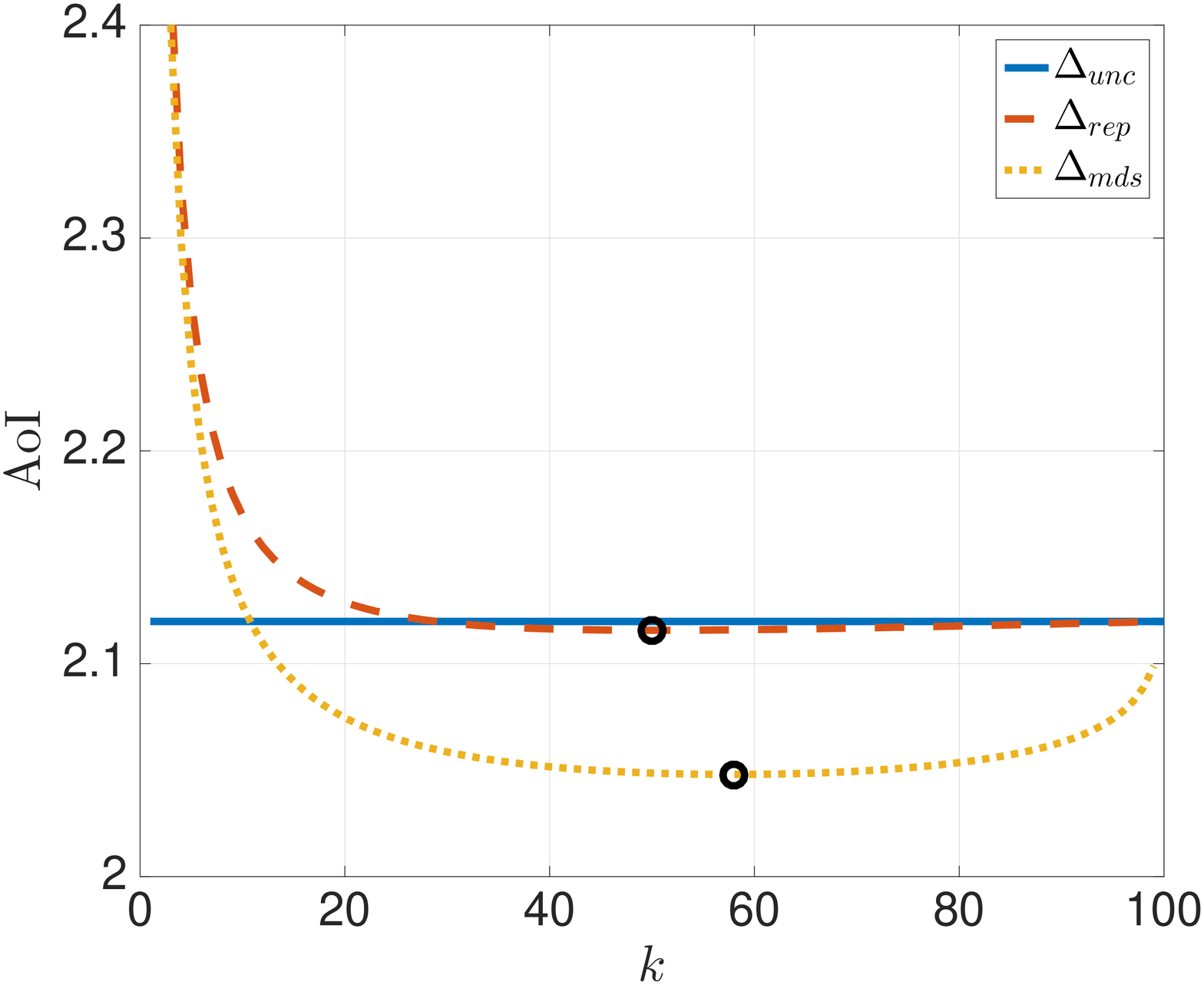}  }
	\caption{ $\Delta_{unc}$, $\Delta_{rep}$ and $\Delta_{mds}$ for varying $k$ with $n=100$, $\lambda =1$, and $c=1$: (a) When $\mu =1 $, (b) when $\mu = 0.5$. Symbol $\circ$ marks the age minimizing optimal $k$ values.}
	\label{fig:mus}
\end{figure}

First, we consider the uncoded, repetition coded and MDS coded schemes. In Figs.~\ref{fig:mus}(a) and~\ref{fig:mus}(b) age performance of the uncoded, repetition coded and MDS coded schemes are presented when $n=100$, $\lambda = 1$, and $c=1$ for $\mu = 1$ and $\mu = 0.5$, respectively, for varying $k$. We observe that in both Figs.~\ref{fig:mus}(a) and~\ref{fig:mus}(b), MDS coded scheme performs the best as expected. Optimal $k$ values for the MDS coded scheme in these cases are $k^* = 69$ and $k^* = 58$, respectively. Moreover, we observe that when $\mu = 1$ optimal $k$ for repetition coded scheme is in fact $k^* = n = 100$. However, when $\mu = 0.5$, we get $k^* = 0.5*100 = 50$. These results are in line with (\ref{repopt}). We also observe in Fig.~\ref{fig:mus}(b) that repetition coded scheme beats the uncoded scheme when $c\mu < 1$. However, as seen in Fig.~\ref{fig:mus}(a) when $c\mu \geq 1$, repetition coded scheme does not present any advantage over the uncoded scheme.

\begin{figure}
	\centering
	\subfloat[\label{fig:optk}]{   \includegraphics[width=.5\columnwidth]{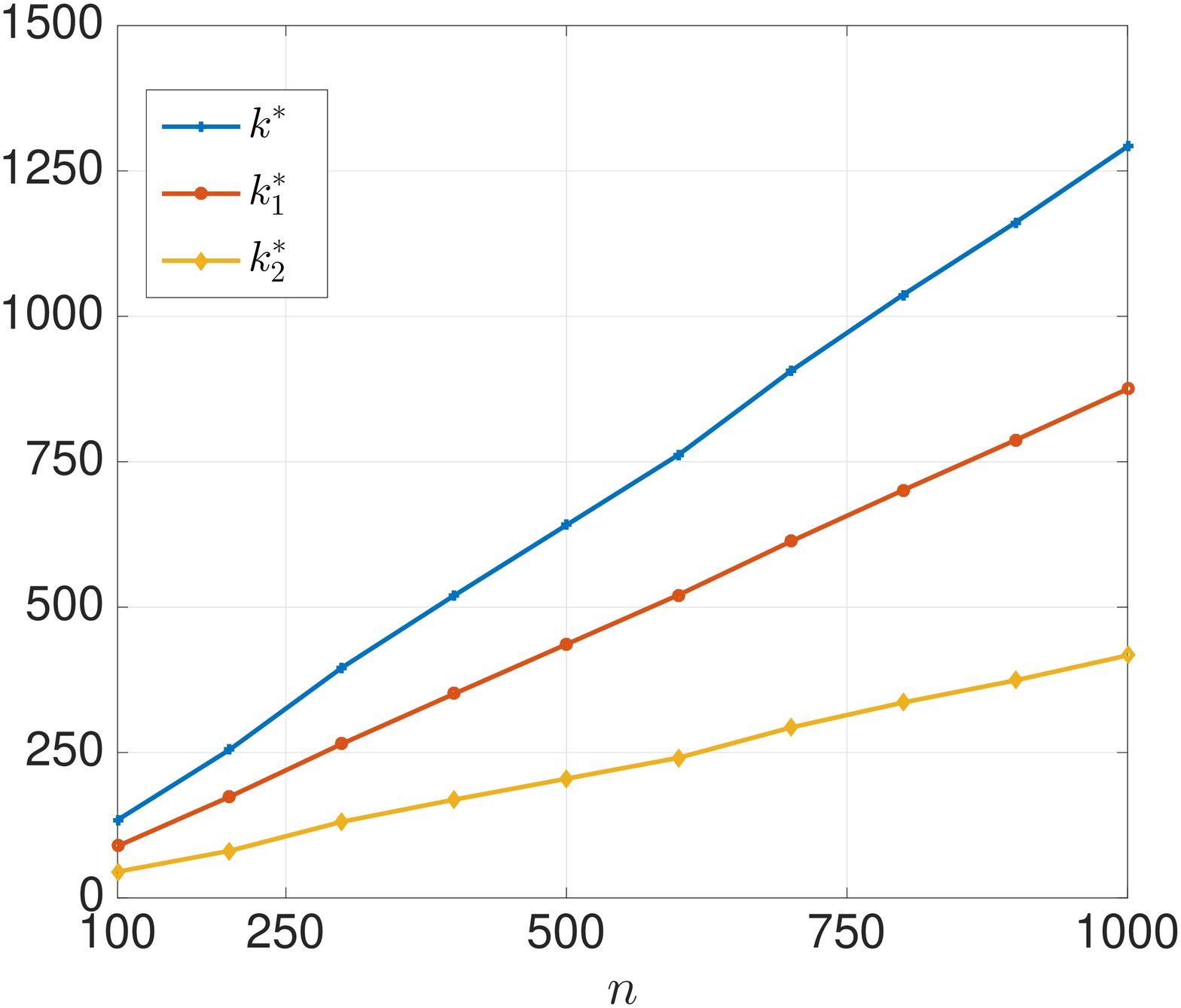}    }%
	\subfloat[\label{fig:sim_mmc}]{  \includegraphics[width=.5\columnwidth]{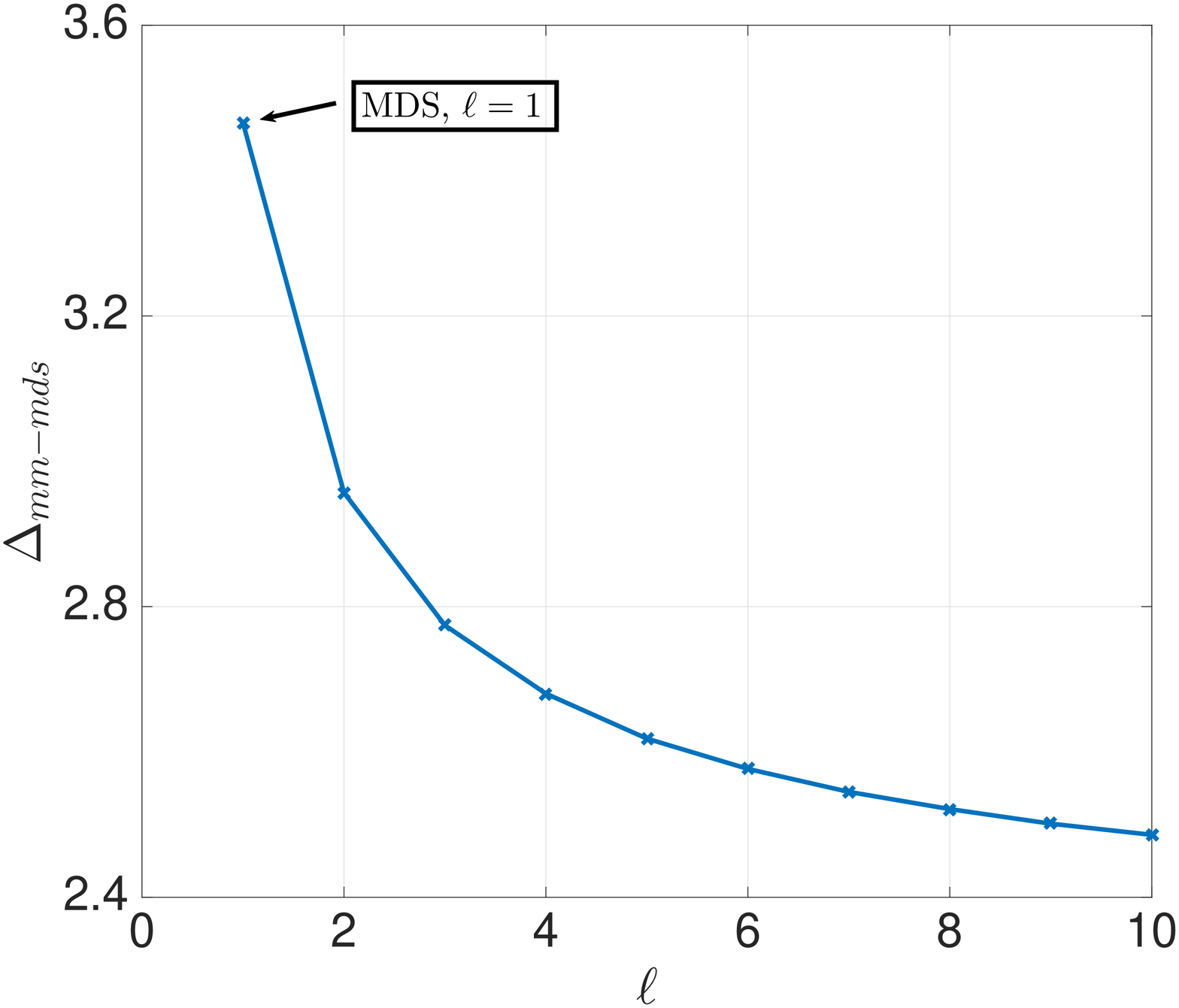}  }
	\caption{ (a) $k^*$, $k^*_1$, and $k^*_2$ values as a function of $n$ for $\ell=2$ with $\mu = 1$, $\lambda=1$, and $c=1$. Note that $k^* = k^*_1 + k^*_2$. (b) Average age, $\Delta_{mm-mds}$, as a function of load $\ell$ when $(n\ell, k^*)$-MDS code is implemented for each $\ell$ for $n=100$, $\mu = 0.01$, $\lambda=1$, and $c=1$.}
	\label{fig:mm-mds}
\end{figure}

Next, we consider the MM-MDS coded scheme. Fig.~\ref{fig:mm-mds}(a) shows the optimal $k$ values as a function of $n$ for $\ell=2$. We observe that as the number of worker nodes, $n$, increases optimal $k$ increases as well. Fig.~\ref{fig:mm-mds}(b) shows the improvement in the average age performance of MDS coded scheme when worker nodes are assigned $\ell$ subpackets to compute with $n=100$, $\mu = 0.01$, $c=1$ and $\lambda=1$. We note that when $\ell = 1$ we recover the performance of the single message MDS coded scheme analyzed in Section~\ref{MDS} and we observe that when multiple subpackets are assigned to each worker, we achieve a lower age than all the other schemes discussed. 

\section{Conclusions}

In contrast to the initial works on age of information which assumed small sized status update packets, we have considered a status update system encountered in emerging data-intensive applications such as UAV and V2V systems, in which the updates are more complex and require processing to extract the useful information. This task is handled by a computation unit consisting of a master node and $n$ worker nodes. We have investigated the age performance of uncoded and coded computation distribution algorithms and showed that the MDS coded task distribution scheme asymptotically outperforms the uncoded and repetition coded schemes. In addition, we observed that assigning multiple computations to each worker node (MM-MDS coded scheme) further improves the age performance of MDS coded scheme. By showing that under certain arrival and service (computation) time profiles minimizing age is equivalent to minimizing the computation time, we have characterized the optimal repetition, MDS and MM-MDS code parameter $k$ (equivalently, $\alpha$s).

\bibliographystyle{unsrt}
\bibliography{IEEEabrv,lib}

\end{document}